\documentclass[manuscript,flushrt]{aastex}
\newcommand{\SSS}{{\it Spitzer}}
\newcommand{\etal}{{et al.}~}

\def\spose#1{\hbox to 0pt{#1\hss}}
\def\simlt{\mathrel{\spose{\lower 3pt\hbox{$\mathchar"218$}}
     \raise 2.0pt\hbox{$\mathchar"13C$}}}
\def\simgt{\mathrel{\spose{\lower 3pt\hbox{$\mathchar"218$}}
     \raise 2.0pt\hbox{$\mathchar"13E$}}}

\slugcomment{2013 October 1 -- ApJS, accepted}

\shorttitle{\SSS-SPT Deep Field Survey}
\begin{document}
\title{THE \SSS-SOUTH POLE TELESCOPE DEEP FIELD: 
SURVEY DESIGN AND IRAC CATALOGS}

\author
{
M.~L.~N.\,Ashby\altaffilmark{1},
S.~A.~Stanford\altaffilmark{2,3},
M.~Brodwin\altaffilmark{4},
A.~H.~Gonzalez\altaffilmark{5},
J.~Martinez-Manso\altaffilmark{5},
J.~G.~Bartlett\altaffilmark{6},
B.~A.~Benson\altaffilmark{7,8},
L.~E.~Bleem\altaffilmark{7,9},
T.~M.~Crawford\altaffilmark{7,10},
A.~Dey\altaffilmark{11},
A.~Dressler\altaffilmark{12},
P.~R.~M.~Eisenhardt\altaffilmark{13},
A.~Galametz\altaffilmark{14},
B.~T.~Jannuzi\altaffilmark{15},
D.~P.~Marrone\altaffilmark{15},
S.~Mei\altaffilmark{16,17,18},
A.~Muzzin\altaffilmark{19},
F.~Pacaud\altaffilmark{20},
M.~Pierre\altaffilmark{21},
D.~Stern\altaffilmark{13},
\and 
J.~D.~Vieira\altaffilmark{22}
}

\altaffiltext{1}{Harvard-Smithsonian Center for Astrophysics, 60 Garden St., Cambridge, MA 02138, USA
[e-mail:  {\tt mashby@cfa.harvard.edu}]}
\altaffiltext{2}{Department of Physics, University of California, One Shields Avenue, Davis, CA 95616, USA}
\altaffiltext{3}{Institute of Geophysics and Planetary Physics, Lawrence Livermore National Laboratory, Livermore, CA 94551, USA}
\altaffiltext{4}{Department of Physics and Astronomy, University of Missouri, Kansas City, MO 64110, USA}
\altaffiltext{5}{Department of Astronomy, University of Florida, Gainesville, FL 32611, USA}
\altaffiltext{6}{AstroParticule et Cosmologie, Universit\'e Paris Diderot,
CNRS/IN2P3, CEA/IRFU, Observatoire de Paris, Sorbonne Paris Cit\'e, 10, rue Alice Domon
et L\'eonie Duquet, 75205 Paris Cedex 13, France}
\altaffiltext{7}{Kavli Institute for Cosmological Physics, University of Chicago, 5640 South Ellis Avenue, Chicago, IL 60637, USA} 
\altaffiltext{8}{Enrico Fermi Institute, University of Chicago, 5640 South Ellis Avenue, Chicago, IL 60637, USA} 
\altaffiltext{9}{Department of Physics, University of Chicago, 5640 South Ellis Avenue, Chicago IL 60637, USA}
\altaffiltext{10}{Department of Astronomy and Astrophysics, University of Chicago, 5640 South Ellis Avenue, Chicago IL 60637, USA}
\altaffiltext{11}{National Optical Astronomy Observatories, 950 N.\ Cherry Ave., Tucson, AZ 85719, USA}
\altaffiltext{12}{The Observatories of the Carnegie Institution for Science, 813 Santa Barbara St., Pasadena, CA 91101, USA}
\altaffiltext{13}{Jet Propulsion Laboratory, California Institute of Technology, 4800 Oak Grove Drive, Mail Stop 169-221 Pasadena, CA 91109, USA}
\altaffiltext{14}{INAF -- Osservatorio di Roma, Via Frascati 33, I-00040, Monteporzio, Italy}
\altaffiltext{15}{Department of Astronomy and Steward Observatory, University of Arizona, 933 N.\ Cherry Ave, Tucson AZ 85719, USA}
\altaffiltext{16}{GEPI, Observatoire de Paris, Section de Meudon, 5 Place J.\ Janssen, 92190 Meudon Cedex, France}
\altaffiltext{17}{Universit\'e Paris Denis Diderot, 75205 Paris Cedex 13, France}
\altaffiltext{18}{Infrared Processing and Analysis Center, Pasadena, CA 91125}
\altaffiltext{19}{Leiden Observatory, Leiden University, PO Box 9513 RA Leiden, The Netherlands}
\altaffiltext{20}{Argelander-Institue f\"ur Astronomie, Auf dem H\"ugel 71, 53121 Bonn, Germany}  
\altaffiltext{21}{Service d'Astrophysique, AIM IRFU/DSM/CEA, 91190 Gif sur Yvette, France}
\altaffiltext{22}{California Institute of Technology, 1200 E.\ California Blvd., Pasadena, CA 91125, USA}

\begin{abstract}

The \SSS-South Pole Telescope Deep Field (SSDF) is a wide-area survey 
using \SSS's Infrared Array Camera (IRAC) to
cover 94\,deg$^2$ of extragalactic sky, making it the largest IRAC survey 
completed to date outside the Milky Way midplane.
The SSDF is centered at ($\alpha,\delta)=($23:30,$-$55:00), in a region that
combines observations spanning a broad wavelength range from numerous 
facilities.  These include millimeter imaging from the South Pole Telescope, 
far-infrared observations from {\sl Herschel}/SPIRE, 
X-ray observations from the {\sl XMM} XXL survey, 
near-infrared observations from the VISTA Hemisphere Survey,
and radio-wavelength imaging from the Australia Telescope Compact Array,
in a panchromatic project designed to address major outstanding questions 
surrounding galaxy clusters and the baryon budget.  Here we describe 
the \SSS/IRAC observations of the SSDF, including the survey design, 
observations, processing, source extraction, and publicly available data products.  
In particular, we present two band-merged catalogs, one for each of the 
two warm IRAC selection bands.  They contain roughly 5.5 and 3.7 million distinct 
sources, the vast majority of which are galaxies, down to the SSDF 
5$\sigma$ sensitivity limits of 19.0 and 18.2\,Vega 
mag (7.0 and 9.4\,$\mu$Jy) at 3.6 and 4.5\,$\mu$m, respectively.
\end{abstract}

\keywords{catalogs --- galaxies: clusters: general --- infrared: galaxies --- surveys }

\section{Introduction}

The large-scale distribution of both baryonic and dark matter and the physical 
laws that govern their evolution are fundamental to cosmology.  Observations 
of the cosmic microwave background constrain the baryon-to-matter ratio to be 
$\sim$16\% (Story \etal\ 2012; Hinshaw \etal\ 2013; Planck Collaboration XVI 
2013).  But observations of galaxy clusters have special advantages for 
understanding the abundance and distribution of matter, because
1) clusters are large enough that they are expected to retain the cosmic 
fraction of baryons, and 
2) clusters are among the few astrophysical objects for which the three dominant 
forms of matter can be observed, namely: the gas mass ($M_{\rm  gas}$), the stellar 
mass ($M_{\rm star}$), and the dark matter mass ($M_{\rm DM}$).  However, 
observations of the matter distribution in clusters as a function of physical scale 
and mass have typically been constrained only at low redshift ($z<0.5$).

At low redshift, the best measurements of the baryon faction to date account 
for $\sim$80\% of the cosmic value (for a recent review, see Kravtsov \etal\ 2009). 
The baryonic mass is dominated by an intra-cluster gas component (Vikhlinin \etal\ 
2006; Arnaud \etal\ 2007; Sun \etal\ 2009), with stars and galaxies typically 
contributing roughly one-tenth as much mass (Gonzalez \etal\ 2007; 
Giodini \etal\ 2009; Lin \etal\ 2012).  However, the stellar mass fraction 
has also been found to be a strong function of cluster mass, with the 
stellar-to-gas fraction decreasing from $\sim$0.25 to 0.05 from groups 
($\sim5 \times 10^{13} M_{\odot}$) to massive clusters ($\sim10^{15} M_{\odot}$).  
This can be interpreted to mean that star formation is much more efficient in the 
lower-mass groups compared to rich clusters (e.g., Muzzin \etal\ 2012).

In contrast with observations, simulations generally predict that the stellar 
fraction is approximately constant with cluster mass, and over-predict the amount 
of stars formed by a factor of $\sim$2--5 in massive clusters 
(e.g., Kravtsov \etal\ 2009).  More recently, simulations have tried to evoke 
various types of astrophysical feedback (e.g., supernovae and active galactic nuclei) 
to suppress star formation and match the observed stellar and intracluster medium 
fraction, with some success (e.g., Planelles \etal\ 2012; Battaglia \etal\ 2013).  

Some of the fundamental questions about galaxy and structure formation
that still await resolution include:\vspace{2.0mm}

1. Why is the baryon fraction in cluster gas less than the universal average?  
% Are the baryons less concentrated than the dark matter?  
Are the missing cluster baryons in stars?  Do the observations 
underestimate the total diffuse stellar mass in the form of intra-cluster light?

2. How do the relationships between cluster gas, dark matter, and stars 
evolve with redshift?  
% Was the relation between M$_{\rm gas}$ and 
% M$_{\rm DM}$ similar or steeper in the past?  
Given that clusters are built hierarchically from mergers of lower-mass groups, 
in the simplest scenario one might expect these relations to be
significantly steeper at higher redshift.

3. What is the radial distribution of the baryonic components as a 
function of M$_{\rm DM}$ and redshift?  

4. What is the assembly history of baryons in the most massive structures?
\vspace{2.0mm}

\SSS's Infrared Array Camera (IRAC; Fazio \etal\ 2004a) 
has already been used to carry out several relatively wide-area extragalactic 
surveys, e.g., SWIRE (Lonsdale \etal 2003), SDWFS (Ashby \etal 2009), and
SERVS (Mauduit et al. 2012), covering more than $70\,$ deg$^2$
all together.  Among many other things, datasets such as these provide an 
efficient means to identify statistical samples of galaxy clusters down to low 
masses even at high redshifts (Eisenhardt \etal\ 2008; Muzzin \etal\ 2009; 
Papovich \etal\ 2010; Stanford \etal\ 2012).  However, the cluster searches
carried out to date are not fully characterized in terms of purity, 
completeness, or mass proxies (via comparison with cluster samples 
identified at other wavelengths), hindering interpretations that depend
upon either cluster mass or ensemble properties.  

\begin{figure}[htb]
\epsscale{0.8}
\plotone{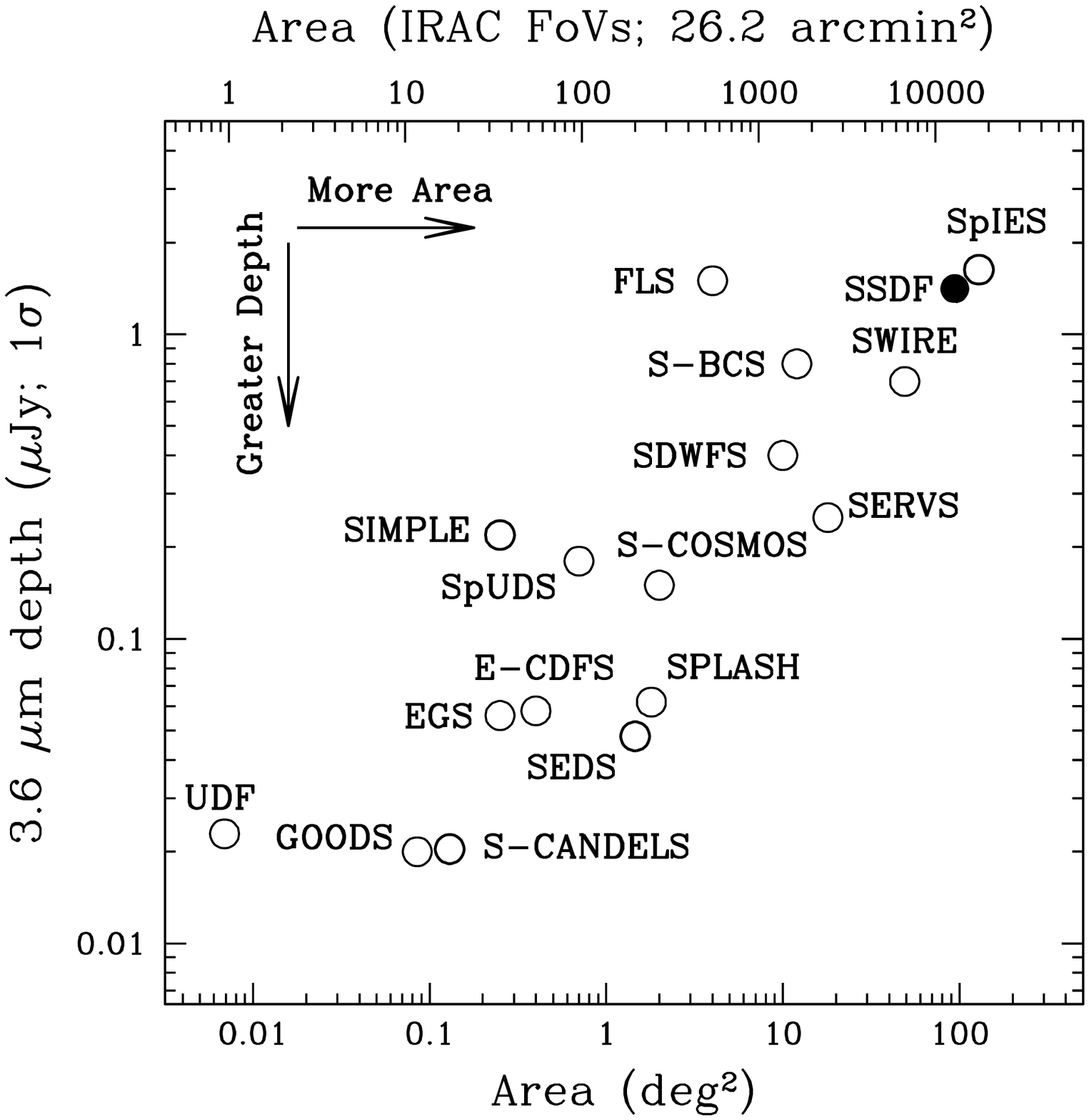}
\caption{Comparison of SSDF 3.6\,\micron\ depth and total area (solid circle) 
to other major \SSS/IRAC extragalactic surveys.   Open symbols indicate 
1$\sigma$ point-source sensitivities
for GOODS (Great Observatories Origins Deep Survey), 
EGS (Extended Groth Strip), E-CDFS (Extended Chandra Deep
Field South), SpUDS (\SSS\ Public Legacy Survey of UKIDSS Ultra-Deep Survey),
SCOSMOS (\SSS\ Deep Survey of {\sl HST} COSMOS 2-Degree ACS Field),
SERVS (\SSS\ Extragalactic Representative Volume Survey),
S-BCS (\SSS-Blanco Cosmology Survey), SWIRE (\SSS\ Wide-area Infrared
Extragalactic Survey), the FLS (\SSS\ First-Look Survey),
SDWFS (\SSS\ Deep, Wide-Field Survey), SEDS (\SSS\ Extended Deep Survey),
S-CANDELS (\SSS-CANDELS), and the \SSS-IRAC Equatorial Survey (SpIES).  
All sensitivities shown are based on low-background estimates made 
with the \SSS\ Sensitivity and Performance Estimation Tool (SENS-PET) 
except for SpIES (which used a high-background estimate), and
the SSDF and SEDS, which are measured from actual data.
\label{fig:etendue}}
\end{figure}

Here we describe a new wide-area IRAC survey that will make it possible to combine
groundbreaking datasets from the South Pole Telescope and {\sl XMM} with coextensive
infrared imaging to address the questions posed above.
We call this the \SSS-South Pole Telescope Deep Field (SSDF) survey.  
The SSDF depth and coverage are compared to those of
other major IRAC survey projects in Figure~\ref{fig:etendue}.
The combination of infrared, millimeter, and X-ray data provides a unique 
opportunity to simultaneously calibrate the IRAC search techniques 
via comparison with Sunyaev-Zel'dovich (SZ) and X-ray cluster samples, and to robustly 
determine the cluster selection functions at the depths simultaneously
probed by the three techniques.  A limitation of the present generation of 
X-ray and SZ surveys is that they lack the sensitivity at high redshift to identify 
clusters down to $10^{14}$ M$_\odot$ and below, as is possible 
with IRAC.  Reaching these mass scales at higher redshifts is 
critical for understanding the assembly history of the present day 
cluster population, as these low-mass systems 
are the direct progenitors of typical present-day clusters, like Virgo.
The IRAC component is essential for finding the
lowest-mass clusters at all redshifts $z > 0.5$ and for determining photometric 
redshifts for complete galaxy samples (Brodwin \etal\ 2006).  It is also
vital for determining galaxy luminosities and sizes within clusters 
(e.g., Brodwin \etal\ 2010), and for estimating the stellar masses of all clusters 
found by any selection technique (e.g., Eisenhardt \etal\ 2008).  
Because the SSDF offers an opportunity to bring the X-ray and millimeter imaging
together with wide-field IRAC data, it is poised to become a uniquely 
valuable resource for galaxy cluster research.  

% More specifically, because the mid-infrared to which IRAC is sensitive is an
% efficient tracer of baryonic mass, our survey will provide:
% 1) a large unbiased mass-selected sample of clusters that spans a significant
% mass range and redshift baseline.  In particular, clusters at $z >$ 1 are important
% because simulations suggest that $> 50$\% of their total dark and baryonic mass
% assembly occurs during that epoch (Ettori \etal\ 2004);
% 2) reliable measures of the total halo mass, total gas mass, and total stellar
% masses for all clusters in the survey.  

% It is anticipated that ultimately the observations presented here will identify
% $\sim$3000 clusters with M $>8\times10^{13}$ M$_{\odot}$ in the redshift range 
% 0.3 $< z <$ 2.0 that
% will be used to make the most complete census of stellar
% baryons in clusters and groups to date.
% The goal is to place strong constraints on cosmological simulations.
% Understanding the distribution of gas and stars on large scales should also
% be important in differentiating between different star formation feedback models,
% currently one of the most debated issues in galaxy formation.

In this contribution, we
describe the \SSS/IRAC survey: the field, the observations, the reduction, and in
particular the resulting catalogs that will serve as the basis for future work
aimed at addressing the unresolved questions surrounding the distribution of
baryons in the Universe.
This paper is organized as follows:  Section~\ref{sec:field} describes the
SSDF field and previous observations relevant to galaxy cluster science.
Section~3 discusses the details of
the SSDF observing strategy and data reduction, and Section~4 describes the
source identification, photometry, and validation.
Section~\ref{sec:catalogs} describes the SSDF catalogs.  In Section~6 we present
preliminary results, including wide-field infrared number counts 
and the infrared color distribution of IRAC-detected galaxies.  
We summarize in Section~7.  Unless otherwise
stated, all magnitudes are given in Vega-relative terms.  Users can convert
to the AB scale by adding 2.792 and 3.265\,mag to the cataloged 3.6 and 4.5\,$\mu$m
magnitudes, respectively.

\section{The {\sl Spitzer}-South Pole Telescope Deep Field}
\label{sec:field}

\begin{figure}[t]
\epsscale{1.0}
\plotone{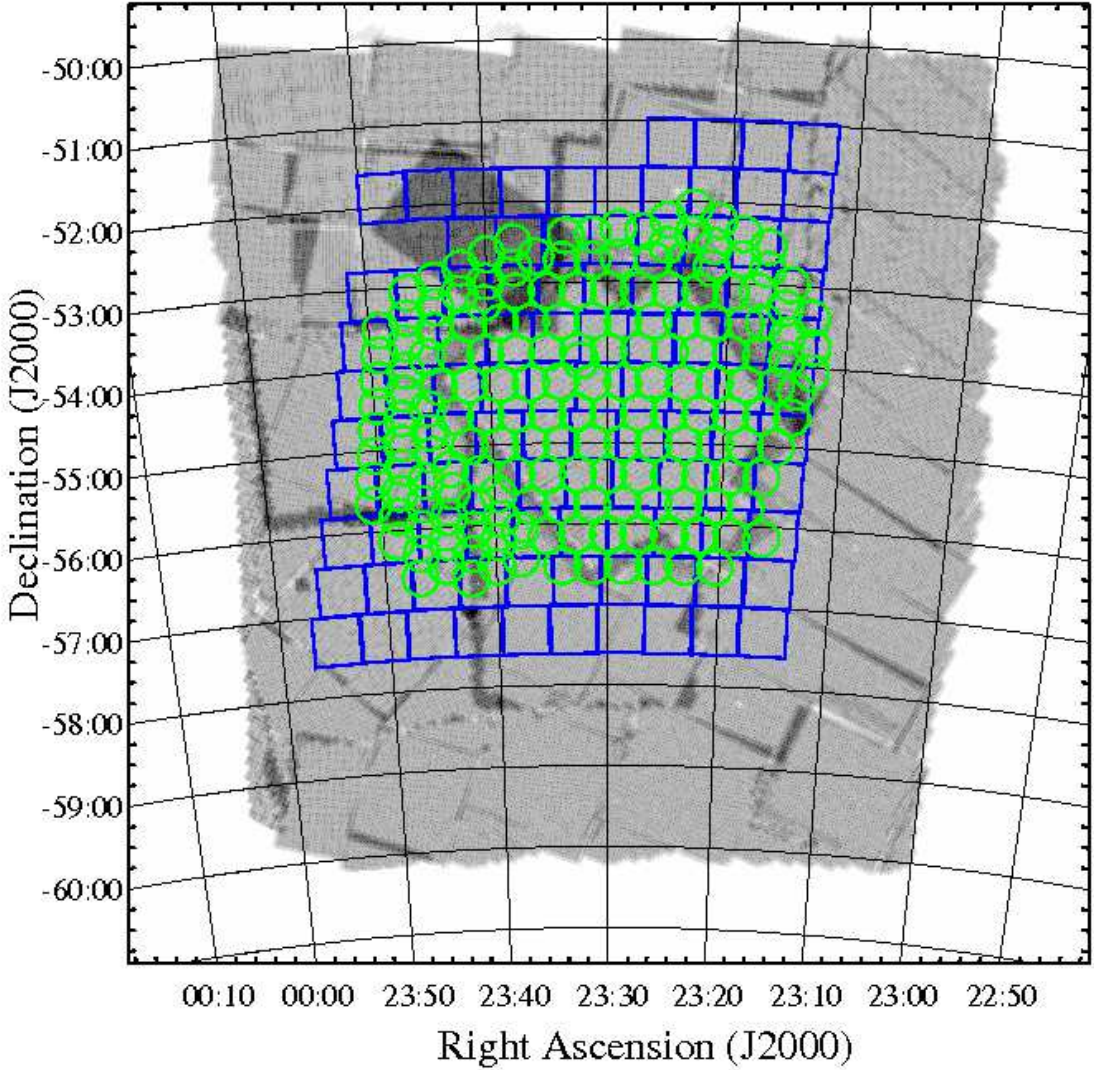} 
% \includegraphics[height=9in,width=7.0in,angle=0]{SSDF_I2_cov_med_reg.pdf}
% \vskip-1.7in
\caption{
The SSDF depth of coverage at 4.5\,$\mu$m (greyscale) including all observations
taken through 2013 February.  The linear stretch ranges from zero (white) to
15 (black).  Most of the field is covered to the designed depth
($4\times30$\,s).  
The coverage at 3.6\,$\mu$m is very similar to that shown here.
Blue squares indicate tiles covered by $griz$ observations from 
the Blanco Cosmology Survey (Desai \etal\ 2012).  Green circles indicate 
{\sl XMM} pointings from the XXL survey, covering a total of 25\,deg$^2$
to a depth of 10\,ks. A full-resolution version
of this figure is available in the version of this article published in
the ApJS.
\label{fig:field2}
}
\end{figure}

\begin{figure}[htb]
\epsscale{1.0}
\plotone{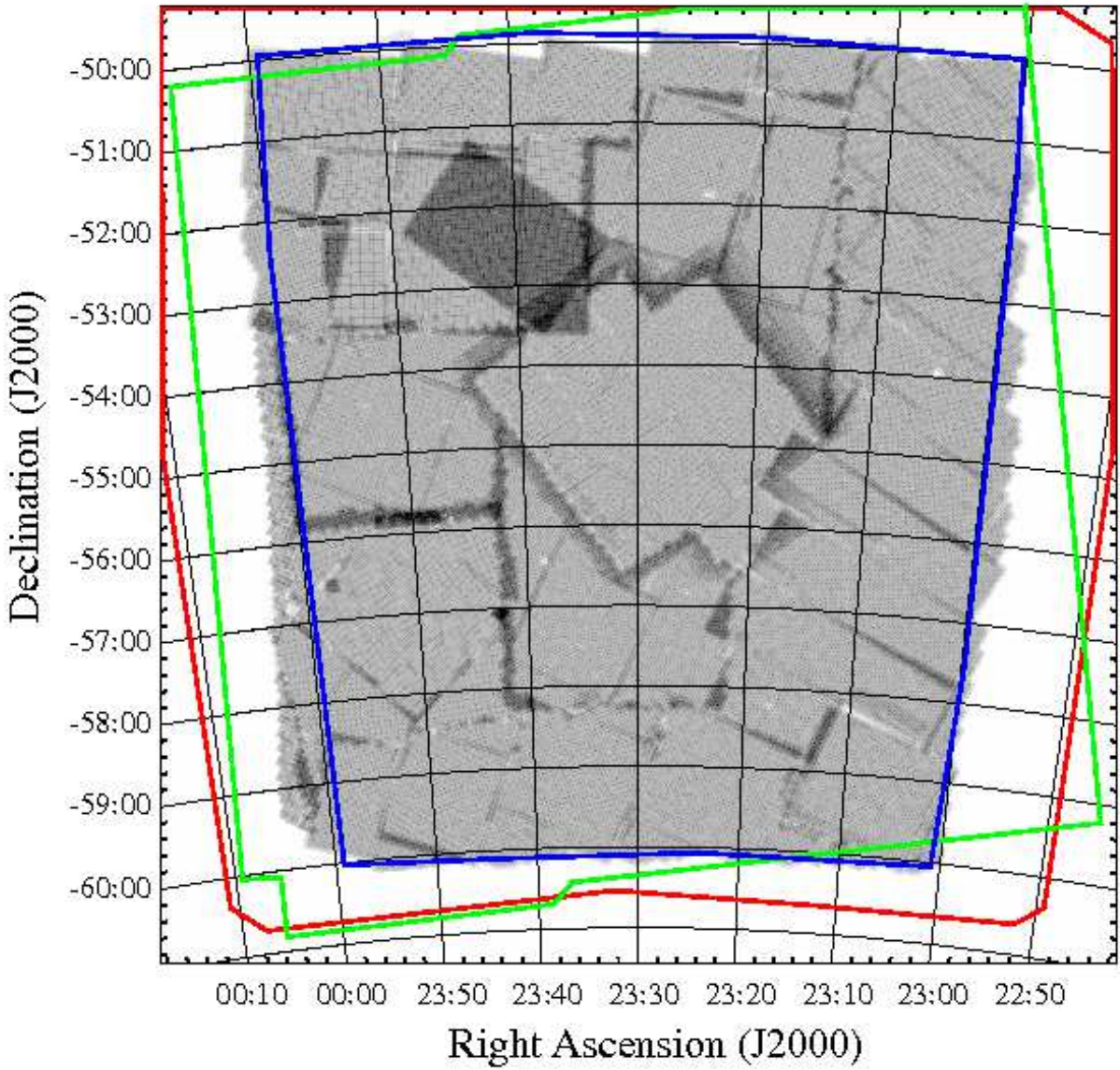}
% \includegraphics[height=9in,width=7.0in,angle=0]{SSDF_ATCA_SPIRE_SPT.pdf}
% \vskip-1.5in
\caption{
As Figure~\ref{fig:field2}, but showing the boundaries of the coextensive 
coverage at long wavelengths.  The extent of coverage obtained via SPT 
imaging at 1.4, 2, and 3\,mm (the SPT deep field) is shown in red, while
the {\sl Herschel}/SPIRE far-infrared imaging is outlined in green, and
the ATCA 16\,cm imaging is outlined in blue.  A full-resolution version
of this figure is available in the version of this article published in
the ApJS.
\label{fig:atca}
}
\end{figure}

We carried out our survey in a field that benefits from 
an abundance of supporting data from X-ray to radio wavelengths, and which
has extremely low levels of Galactic dust emission, being among
the cleanest 1\% of contiguous 100\,deg$^2$ regions on the sky
as measured in the 100\,$\mu$m IRAS map (Finkbeiner et al. 1999).
The SSDF is centered at $(\alpha,\delta)=(23$:30,$-$55:00).
The relationship of the SSDF coverage to that of coincident surveys 
at other wavelengths is shown in Figures~\ref{fig:field2} and \ref{fig:atca},
is summarized in Table~\ref{tab:ancillary}, and is described in detail below.

Their sensitivity to baryons in the intracluster
medium makes X-ray observations 
particularly important for the study of galaxy clusters, and the {\sl XMM} 
mission carried out a large-area survey specifically designed to be very 
sensitive to clusters.  This is the XXL survey (Pierre \etal\ 2011), 
of which a 25\,deg$^2$ portion is located in the SSDF.  The XXL survey 
was performed in 2011--2013 with 10\,ks integrations per pointing,
and the resulting cluster catalogs are now being confirmed via 
spectroscopic followup programs at the AAT, NTT, and VLT 
(M. Pierre et al., in prep.)  
The XXL survey is the largest deep-wide X-ray survey for galaxy 
clusters in existence, with an area five times greater than existing {\sl XMM} surveys 
and sensitivity sufficient to study clusters beyond $z=1$ down to masses
of $\sim10^{14}$ M$_\odot$.  The cross-comparision between the infrared
and X-ray cluster catalogs will provide unique information 
about the purity and completeness of the samples as well as on the physics
of high-redshift clusters.

The entire SSDF was also imaged at far-infrared wavelengths in 2011 
with {\sl Herschel}'s Spectral and Photometric Imaging REceiver 
(SPIRE; Griffin \etal\ 2010).  
The SPIRE imaging reaches 1$\sigma$ depths of roughly 
10\,mJy\,beam$^{-1}$ at 250, 350, and 500\,$\mu$m.  The SPIRE data will be 
very useful for studies of submm galaxies and for cross-correlation studies
involving Cosmic Microwave Background lensing maps (e.g., Holder \etal\ 2013),
and should probe ULIRG activity in galaxy clusters.  
Shallow infrared imaging is already available from the 
all-sky {\sl WISE} survey (Wright \etal\ 2010), 
which reaches 1 and 6\,mJy at 12 and 22\,$\mu$m, respectively.  
In the radio, the full field is being imaged with the Australia Telescope
Compact Array (ATCA) at 16\,cm down to 40\,$\mu$Jy\,beam$^{-1}$ with a 
7\arcsec\ beam.

In the visible/near-infrared regime, the SSDF already has extensive
coverage from both completed and ongoing surveys.  
The Blanco Cosmology Survey (Desai \etal\ 2012;
L.\ Bleem \etal\ in preparation) surveyed $\sim$30 deg$^2$ of the SSDF 
in the $griz$ bands down to roughly 23\,AB mag (10 $\sigma$).
The ongoing Dark Energy Survey will obtain $griz$ imaging
down to approximately 
24\,AB mag (10$\sigma$) over a large area that includes the entire SSDF.
% DES data are to be made public one year after acquisition.
In the near-infrared, Data Release 1 of the VISTA Hemisphere Survey 
(McMahon \etal, in preparation) has already
covered nearly the entire field in $JHK$ to 5$\sigma$ limits of
21.2, 20.8, and 20.2\,AB mag. 

Roughly 12\,deg$^2$ in the center of the SSDF were observed with 
\SSS/IRAC in Cycle 4 (PI Stanford, PID 40370) with three
dithered 30\,s exposures.  These observations were incorporated into
the SSDF and are hereafter referred to as the S-BCS (\SSS-Blanco Cosmology 
Survey).  Because it was obtained during \SSS's cryogenic mission,
the S-BCS dataset includes exposures at 5.8 and 8.0\,$\mu$m, although
those two long-wavelenth bands were not analyzed as part of the 
present work because they are significantly less sensitive than
our new IRAC imaging.

The SSDF has already been fully covered at 95, 150, and 220\,GHz 
to 1$\sigma$ depths of 2, 1, and 3\,mJy\,beam$^{-1}$, respectively
by the South Pole Telescope (SPT) during  
its survey of more than 2500\,deg$^2$ of 
the southern sky (Carlstrom et al.\ 2011; Story et al.\ 2012).
The SPT provides an efficient means to
detect high-redshift dusty galaxies (e.g., Vieira \etal 2013), 
but it is particularly effective for identifying 
galaxy clusters via the Sunyaev-Zel'dovich effect (SZ; 
Sunyaev \& Zel'dovich 1972).  The SZ selection technique 
is relatively insensitive to cluster redshift, so the SPT observations
are capable of identifying galaxy clusters out to great distances
(e.g., Reichardt \etal\ 2013).  Over the next four years, even deeper 
observations by SPTpol (Austermann \etal\ 2012), which has
already imaged this field to nearly twice the depth of the
larger SPT survey (Story et al.\ 2012) will further improve the 
quality and depth of the millimeter data in this field.  
Its southern location means 
that it is well-placed for followup observations with the 
Atacama Large Millimeter/Submillimeter Array, and 
is a promising candidate for selection as one of the
Large Synoptic Survey Telescope deep drilling fields.

\section{Observations and Data Reduction}

\subsection{Mapping Strategy}

The observing plan drew heavily on our team's prior experience with
the IRAC Shallow Survey (Eisenhardt \etal\ 2004).  The observations 
were arranged to account for and smoothly extend the pre-existing IRAC coverage 
from the S-BCS (3$\times$30\,s depth over 12 deg$^2$).  They also
had to accommodate the position angle (PA) 
constraints of the two observing windows opening six months apart
as well as the SSDF boundaries.  The SSDF lies between Declinations
$-50$\,deg and $-60$\,deg, and from Right Ascension
23\,hr to slightly east of 0\,hr (Figure~\ref{fig:field2}).
To accommodate these constraints, the SSDF 
was covered by Astronomical Observation Requests (AORs) having coverage 
footprints of various (sometimes irregular) shapes and sizes.  When possible, 
standard IRAC grids were used to cover $\sim1\times$1\,deg$^2$ areas in four passes,
each pass obtaining a single 30\,s exposure per position.   Around the edges of and in
between these grids, we designed AORs in fixed cluster mode so as to
optimally cover irregular areas.  Like the AORs covering areas having
more regular shapes, these gerrymandered observations 
were also organized into groups of four single-pass AORs to cover
each area, with a single 30\,s exposure obtained in each
pass.  In any one area, the four AORs were obtained over the course
of a two-day window.  The goal was to achieve 
uniform coverage to the greatest extent possible. 

% With prior knowledge of the PAs of the tiles, 
% we were able to design Astronomical Observation Requests (AORs) 
% that mosaiced the full SSDF without large coverage gaps (but see below).
% The tiles were mapped with sets of single-pass mapping AORs each
% utilizing a 30\,s integration time.  
% Three from each set of four AORs executed standard IRAC grids, typically
% $12 \times 12$ pointings.  The fourth AOR was done in fixed
% cluster offset mode, and was gerrymandered to optimally cover the
% survey area with minimal overlap of adjacent tiles.

Each AOR required 1--6\,hr to execute, ensuring at least 1\,hr 
between successive observations of each sky position.  For typical 
asteroid motions of 25\arcsec\,hr$^{-1}$, asteroids will move distances
much greater than the IRAC point spread function between AORs.
The four observations of each sky position therefore
allowed asteroids to be excluded from the final mosaics with standard outlier
rejection techniques.

Our mapping strategy was to dither the exposures on
% incorporated several elements to facilitate
% self-calibration of the data by maximizing inter-pixel correlations
% (e.g., Arendt, Fixen, \& Moseley 2000).  We 
% {\em dithered} the observations on
small scales, and offset by one-third of an IRAC field of
view between successive passes through each AOR group. This provided
inter-pixel correlation information on both small and large scales.
% Together, these mapping strategies allowed for significantly
% enhanced self-calibration of the data.  
Our observing strategy is therefore
very robust against bad rows/columns, large scale cosmetic defects
on the array, after-images resulting from saturation due to bright
stars, variations in the bias level, and the color dependence of
the IRAC flat-field across the array (Quijada \etal\ 2004).

Although the four-AOR observations of specific areas were performed
consecutively, spacecraft visibility constraints meant that coverage
of the full SSDF had to be accumulated in separate campaigns
spaced roughly six months apart.  These took place in 2011 July-August,
2012 January-February, 2012 July-September, and 2012 December-2013 February.   
Because adjacent regions were inevitably observed at slightly
different PAs, obvious and irregular coverage gaps were evident between
adjacent mapping AOR sets as the observations accumulated.  
These were covered in the latter two epochs with
cluster-mode AORs during which the exposures were placed
to fill the coverage gaps as smoothly as possible given the
spacecraft visibility limits and position angles.  Like the AORs executed
earlier, these irregular cluster-mode AORs were done in sets
of single-pass rasters intended to accumulate a total of 
four dithered 30\,s exposures everywhere.

\subsection{Data Reduction}
\label{sec:reduction}

To the maximum extent possible, identical reduction procedures were 
applied to all SSDF and S-BCS data so as to ensure uniform data quality 
throughout the field.
The data reduction was based on version S18.18.0 of the IRAC Corrected 
Basic Calibrated Data (cBCD) exposures for the first SSDF campaign;
version S19.1.0 was used for all other, later campaigns.  
The cBCD data were used because some of the salient instrument
artifacts (e.g., multiplexer bleed) are automatically corrected
by the cBCD pipeline.  Other artifacts (e.g., scattered light) 
are flagged in the cBCD pipeline-adjusted pixel masks for each frame.
Both features of the cBCD data make them optimal for our purposes.

To remove
slowly-decaying residual images from unrelated observations of 
bright objects, all 3.6 and 4.5\,$\mu$m 
cBCD frames were object-masked and median-stacked on a per-AOR basis. 
The resulting stacked images (presumed to represent blank sky) were 
visually inspected and subtracted from individual 
cBCDs within each AOR.  This created sky-subtracted versions of the
cBCDs that were free of long-term residual images 
arising from prior observations of bright sources.
Residual images with short decay times arising from observations
of bright stars during the SSDF observations themselves were not
addressed by this method, however.  Pixels flagged as potentially
contaminated with such residuals by the IRAC pipeline were masked.
The sky-subtracted cBCDs were then examined individually and processed
using custom software routines to correct column-pulldown 
effects associated with bright sources.  The
code, known as the ``Warm-Mission Column Pulldown Corrector,"
is publicly available at the \SSS\ Science Center\footnote{\tt
http://ssc.spitzer.caltech.edu/archanaly/contributed/browse.html}.

After these preliminaries, the SSDF exposures and the coincident
IRAC imaging from the S-BCS were mosaiced with {\tt IRACproc} 
(Schuster \etal 2006).  
{\tt IRACproc} augments the capabilities of the standard IRAC 
reduction software (MOPEX) by calculating the spatial derivative 
for each image pixel and adjusting the clipping algorithm accordingly.  
Pixels where the derivative is low (in the field) are clipped 
more aggressively than are pixels where the spatial derivative 
is high (point sources).  This avoids downward biasing of point source 
fluxes in the output mosaics that might otherwise occur because 
of the slightly undersampled IRAC point spread function (PSF).
The software was configured to automatically flag and reject cosmic 
ray hits based on pipeline-generated masks together with the adjusted 
sigma-clipping algorithm for spatially coincident pixels.

\begin{figure}[htb]
\epsscale{1.0}
\plotone{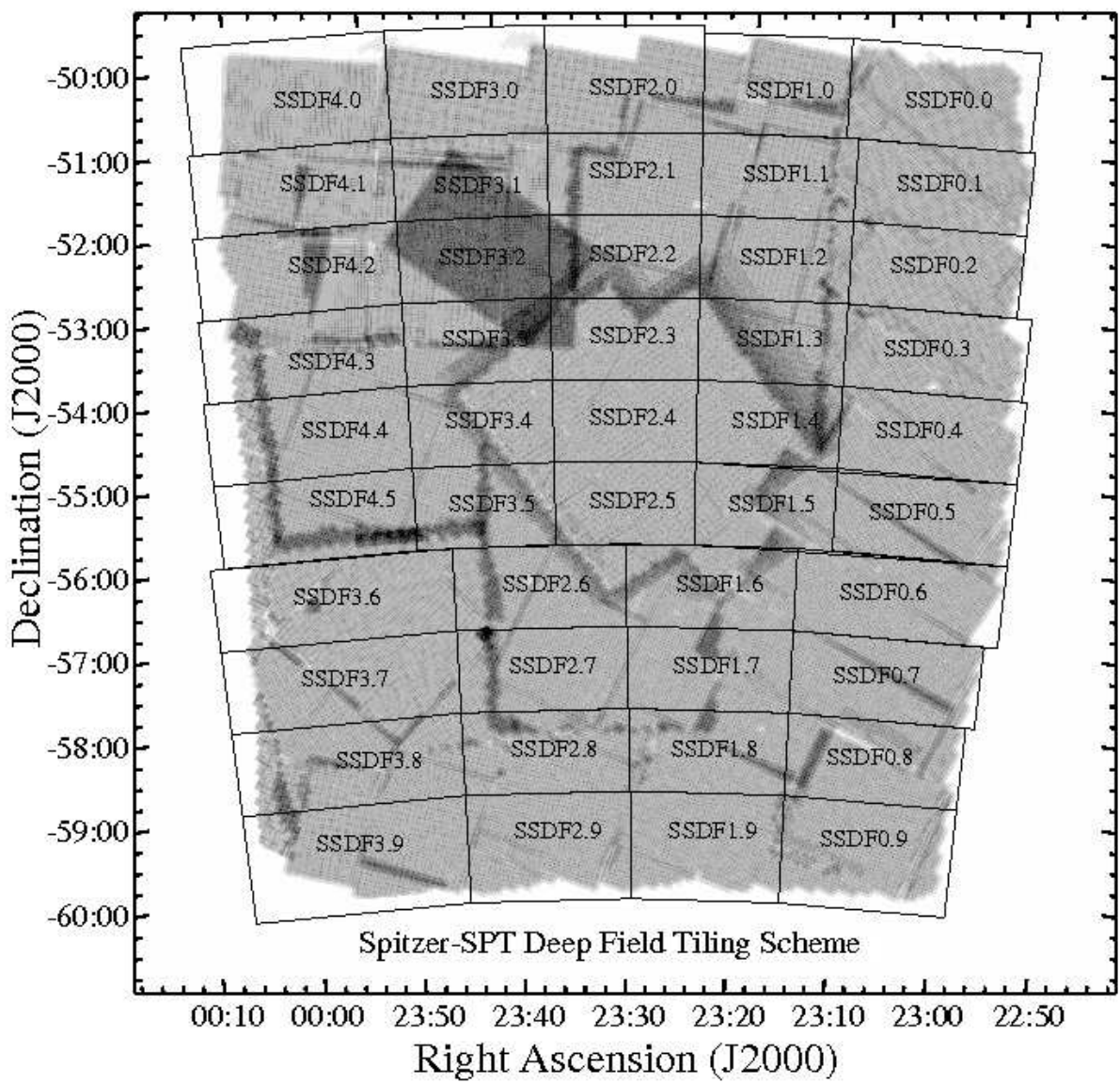}
% \includegraphics[height=9.0in,width=7in,angle=0]{SSDF_tiles.pdf}
% \vskip-1.5in
\caption{ 
As Figure~\ref{fig:field2} but showing the tiling scheme 
used to define sub-regions of the SSDF in which
mosaics were constructed as described in Section~\ref{sec:reduction}.  
The dark rectangle centered in tile SSDF3.2 
at roughly ($\alpha,\delta)=($23:40,$-$52:30) 
is covered to $8\times30$\,s, twice the nominal SSDF depth.  
The original $3\times30$\,s S-BCS observations (now
covered to $4\times30$\,s depth) are apparent as
the irregular region of uniform coverage centered on tile SSDF2.4.
A full-resolution version
of this figure is available in the version of this article published in
the ApJS.
\label{fig:field}
}
\end{figure}

The IRAC mosaics were organized 
into pairs of coextensive tiles each covering roughly $2\times1$\,deg$^2$ 
sub-fields at both 3.6 and 4.5\,$\mu$m.
A total of 46 tiles were required to cover the full SSDF.  The IRAC
coverage and the tile dimensions and locations are defined in 
Figure~\ref{fig:field} and Table~\ref{tab:tiles}.  
By construction, the World Coordinate Systems of all SSDF tile 
pairs are tied to the coordinates of objects in the 
Two Micron All Sky Survey Point Source Catalog 
(2MASS, Skrutskie \etal\ 2006), and are in 
perfect pixel-by-pixel registration.

The resulting 92 final mosaics/coverage map pairs (one pair per tile per band),
are publicly available from the Exploration Science Programs webpage at 
the Spitzer Science 
Center\footnote{\tt http://irsa.ipac.caltech.edu/data/SPITZER/docs/spitzermission/observingprograms/es/}.  All SSDF mosaics were built with 0\farcs6 pixels and
have units of MJy\,sr$^{-1}$.

\section{Source Extraction and Photometry}
\subsection{Source Identification}
\label{sec:ids}

We detected and photometered sources in the SSDF mosaics
with SExtractor (ver. 2.8.6; Bertin \& Arnouts 1996).  SExtractor
is well-suited to the relatively sparse SSDF mosaics, where there are
numerous source-free pixels available for robust sky background estimation.
We adopted the SExtractor parameter settings shown in Table~\ref{tab:settings}.  
The coverage maps constructed by {\tt IRACproc} were used as detection weight 
images.  Custom flag images were constructed to identify and exclude all 
mosaic pixels covered by fewer than two exposures.

Photometry was measured within nine apertures having diameters
ranging from 2\arcsec\ to 10\arcsec\ in 1\arcsec\ increments, plus two additional
apertures of diameter 15\arcsec\ and 24\arcsec, or eleven apertures in all.
By comparing photometry measured in the 24\arcsec\ aperture (i.e., the aperture 
adopted by the Instrument Team as the fiducial photometric aperture for IRAC) 
to that in all other
apertures for well-detected, isolated sources, we obtained empirical
estimates of the aperture corrections, which are given in 
Table~\ref{tab:apcorr}.  We also obtained
MAG\_AUTO magnitudes, for which SExtractor measures fluxes
interior to elliptical apertures having sizes and orientations
determined using the second-order moments of the light distribution
measured above the isophotal threshold.  We compared the MAG\_AUTO
photometry to the corrected-to-total aperture magnitudes and found
that the MAG\_AUTO estimates were systematically fainter by roughly 0.05\,mag.
In other words, the corrected aperture magnitudes are consistent
with each other, but the MAG\_AUTO measurements are 0.05\,mag fainter
on average. 

We used SExtractor in dual-image mode.
In this configuration, sources are detected, their
centers are located, and their apertures are defined in one image,
and subsequently photometry is carried out in another image using
those pre-established apertures and source centroids.   
The dual-image approach forces SExtractor to measure 
the emission from all sources over identical areas in both bands, 
ensuring that the resulting photometry will yield accurate source colors.
SExtractor was configured to define a source as a set of four 
or more connected pixels each lying 0.5$\sigma$ above the estimated 
background.  We first used the 4.5\,$\mu$m mosaics as the detection images;
both the 3.6 and 4.5\,$\mu$m mosaics were used in turn as the photometry
images.   The process was followed in all 46 tiles covering the SSDF, 
resulting in 46 pairs of single-band catalogs.

Because the source extraction was performed in dual-image mode, the
separate 3.6 and 4.5\,$\mu$m SExtractor catalog pairs generated
for each of the two selection bands were in line-by-line 
registration.  We generated band-merged catalogs
by combining photometry from these catalog pairs for all SSDF tiles.
The band-merged, tile-based catalogs were trimmed to exclude overlapping 
regions at the tile boundaries given in Table~\ref{tab:tiles},
and all aperture photometry was corrected to total magnitudes using
the empirical aperture corrections from Table~\ref{tab:apcorr}.  
A full-field catalog covering the entire SSDF was then created by combining
the trimmed, band-berged, aperture-corrected photometry from all tiles.
We chose to include the corrected 4\arcsec\ and 6\arcsec\ diameter aperture
magnitudes in the final catalogs, along with the SExtractor MAG\_AUTO
magnitudes.  

Finally, the above process was repeated using the 3.6\,$\mu$m mosaics
as the detection images.  The result is a pair of full-field band-merged 
SSDF catalogs -- one selected at 3.6\,$\mu$m, and another selected 
at 4.5\,$\mu$m.

\subsection{Survey Depth, Completeness, and Astrometric Reliability}
\label{sec:completeness}

\subsubsection{Survey Depth and Completeness Estimation}

We used a Monte Carlo approach to estimate the survey completeness and
sensitivity by placing numerous simulated sources in the SSDF mosaics at 
random locations and then photometering them in an identical manner to
that used for the original mosaics.  Specifically, we inserted 
simulated objects in both the 3.6 and 4.5\,$\mu$m mosaics for five different 
SSDF tiles (labeled SSDF1.6, 2.4, 2.6, 3.2, and 3.5 in Figure~\ref{fig:field}) 
chosen as representative of the range of IRAC depths of coverage that we obtained. 
The total area in which the simulated sources were inserted
and subsequently photometered therefore samples roughly 10.6\,deg$^2$ of the
total survey field.

The simulated sources were randomly assigned magnitudes between 
10 and 21\,Vega mag.  Hundreds of simulated sources in this range 
were simultaneously placed at random locations 
in each of the five tiles employed for this purpose.  The number of simulated 
sources inserted at one time was restricted to a small percentage of the 
total number of objects apparent in the field, so as to avoid 
artificially-induced source confusion effects.  Nonetheless, because
the simulated sources were allowed to fall anywhere in their respective
tiles, including atop real sources, the simulations do account realistically
for the effects of source confusion.
The process was iterated, so that a total population of 20,000 simulated
sources was ultimately analyzed in each of the 0.5\,mag-wide bins
we constructed to span the magnitude range we considered.

\begin{figure}[htb]
\epsscale{1.0}
\plotone{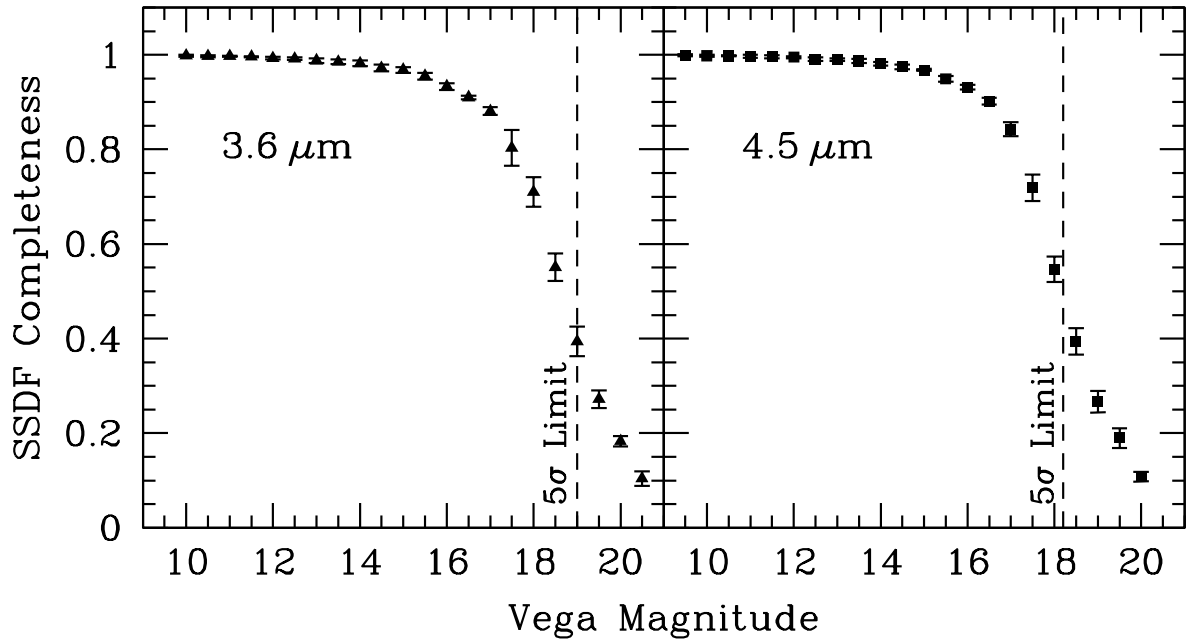}
% \includegraphics[height=8.5in,width=7in,angle=0]{SSDF_complete_new.pdf}
% \vskip-3in
\caption{
Recovered fraction of simulated SSDF sources as
a function of input magnitude, based on the simulations
described in Section~\ref{sec:completeness}.  
\label{fig:comp}
}
\end{figure}

After processing the modified mosaics with SExtractor in exactly the same way 
as was done for the original mosaics, the resulting catalogs were compared
to determine the completeness as a function of magnitude.
The comparison was performed in catalog space with MAG\_AUTO magnitudes 
using a simple
position-matching criterion.  An additional constraint was imposed, 
requiring that a valid detection of a simulated source had to 
yield a measured magnitude that fell within 0.5\,mag of 
its {\sl a priori} known magnitude to account for source confusion.
This procedure was repeated in
both bands for all five tiles tested.  The results are given in 
Table~\ref{tab:comp} and shown in Figure~\ref{fig:comp}.

The SSDF catalogs include only sources with aperture-corrected (total) 
magnitudes brighter than the achieved sensitivity levels in at least 
one of the two SSDF bands.  We defined the SSDF sensitivity as the 
magnitude at which the empirical uncertainty in the SExtractor-estimated 
fluxes reached approximately the 5$\sigma$ level (0.2\,mag).  
This occurred at [3.6]=19.0\,mag (7.0\,$\mu$Jy) and [4.5]=18.2\,mag (9.4\,$\mu$Jy) 
for 4\arcsec\ diameter apertures, similar to the sensitivity
reported by Eisenhardt \etal\ (2004) for the first IRAC
survey of Bo\"otes: [3.6]=19.1\,mag and [4.5]=18.3\,mag.
The SSDF thus achieves 5$\sigma$ sensitivities similar to those predicted 
by the online Sensitivity Performance Estimation Tool (SENS-PET):
[3.6]=19.2\,mag and [4.5]=18.6\,mag assuming low background conditions.

\subsubsection{Astrometric Reliability}
\label{sec:astrometry}

\begin{figure}[htb!]
\epsscale{1.0}
\plotone{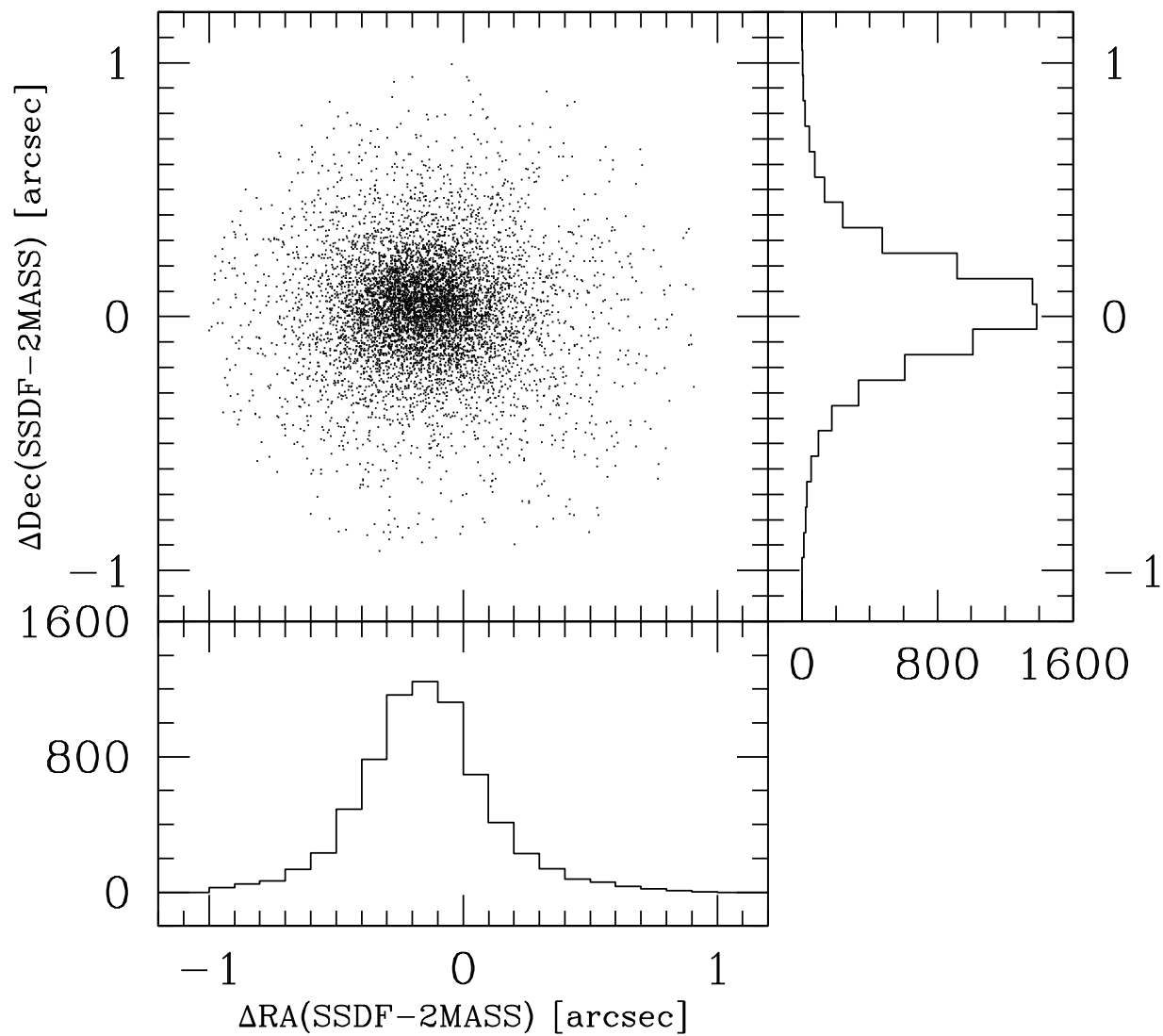}
% \epsscale{1.0}
% \includegraphics[height=8in,width=7in,angle=0]{SSDF_2MASS.pdf}
\caption{
Coordinate offsets measured for IRAC-detected sources also
detected in the 2MASS Point Source Catalog.  The mean differences are 
$-0\farcs15\pm0\farcs26$ in Right Ascension and $0\farcs03\pm0\farcs23$ 
in Declination, giving a total radial uncertainty of just 
$0\farcs15$ (1$\sigma$) relative to 2MASS, or less than
one-tenth of the FWHM of the IRAC point spread function at 3.6\,$\mu$m.
\label{fig:2mass}
}
\end{figure}

To estimate the accuracy of the SSDF astrometry, we compared SSDF IRAC
positions of bright but unsaturated sources to those in the 2MASS 
Point Source Catalog (Skrutskie \etal\ 2006).  
We performed a search within 1\arcsec\ of the positions of IRAC sources 
to identify their 2MASS counterparts.  The distributions of coordinate 
offsets for the 3.6 and 4.5\,$\mu$m sources are shown in 
Figure~\ref{fig:2mass}. The astrometric 
discrepancies are small compared to the size of a SSDF pixel: 
the mean difference (SSDF$-$2MASS) was just $-0\farcs15\pm0\farcs26$ in 
Right Ascension and $0\farcs03\pm0\farcs23$ in Declination.  
The total radial uncertainties are therefore $0\farcs15$ 
(1$\sigma$) relative to 2MASS.  This is about one-fourth of an 
SSDF mosaic pixel, and less than one-tenth of the full width at half
maximum of the IRAC PSF in either band.
This is comparable to the astrometric precision obtained in other \SSS/IRAC
surveys, e.g., SDWFS (Ashby \etal 2009).

\section{SSDF Catalogs}
\label{sec:catalogs}

Both versions of the band-merged SSDF catalogs are presented here,
one for each of the two selection bands (3.6 and 4.5\,$\mu$m).   
The catalogs contain a total of $5.5\times10^6$ and $3.7\times10^6$ 
3.6\,$\mu$m- and 4.5\,$\mu$m-selected sources, respectively,
down to the 5$\sigma$ detection thresholds.  The formats of the two
catalogs are identical and are defined in Table~\ref{tab:catalog}.
All aperture magnitudes in both catalogs have
been corrected to total magnitudes using the aperture corrections given
in Table~\ref{tab:apcorr}.  The catalogs also contain MAG\_AUTO total
magnitude estimates.  All photometric estimates are provided with 
1$\sigma$ uncertainty estimates generated by SExtractor, 
and are also given as flux densities in units of $\mu$Jy.  
In addition to the photometry, the SSDF catalog provides a number of 
SExtractor-derived descriptors for each source as measured in the 
selection band, 3.6 or 4.5\,$\mu$m as appropriate (Table~\ref{tab:catalog}).

\section{Discussion}
\label{sec:discussion}

% \subsection{IRAC Color Distribution}

\begin{figure}[htb]
\epsscale{1.0}
\plotone{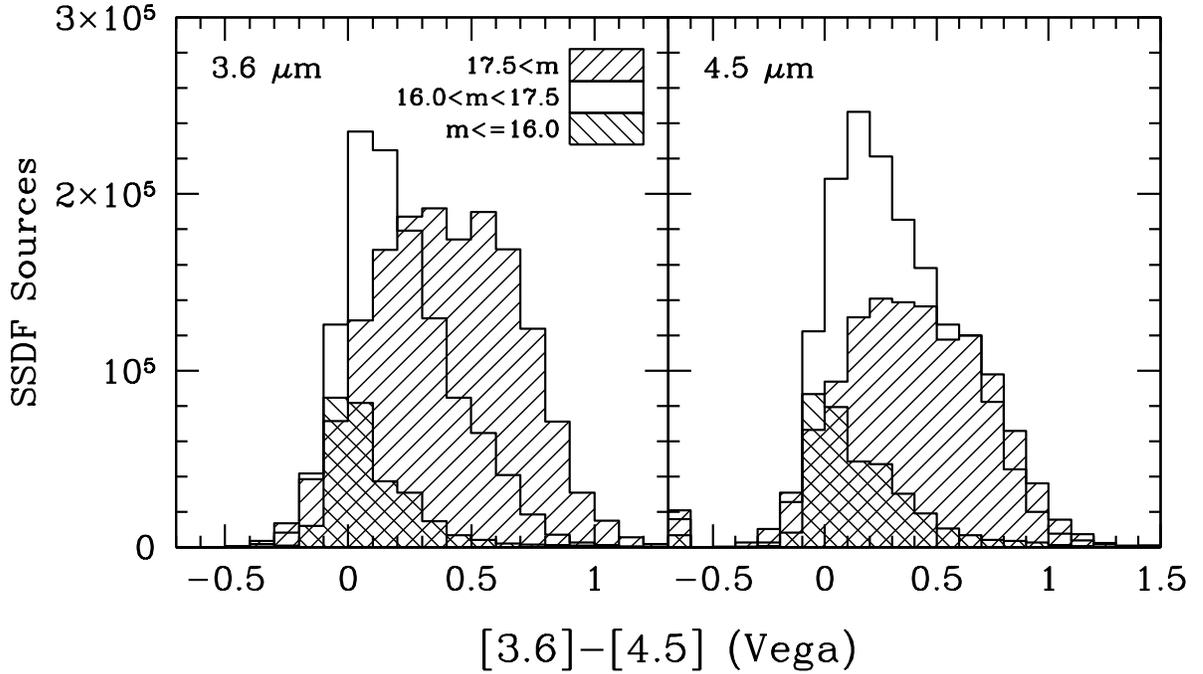}
% \includegraphics[height=8in,width=7in,angle=0]{SSDF_mhist.v9.pdf}
% \vskip-2in
\caption{
The color distributions of SSDF sources detected with greater than
$5\sigma$ significance in both IRAC bands.  All colors are measured 
from aperture-corrected 4\arcsec-diameter aperture photometry.
{\sl Left panel:}  Colors for SSDF sources selected at 3.6\,$\mu$m.
The larger hatched histogram corresponds
to the faintest sources, i.e., those fainter than $[3.6]=17.5$\,mag.
The brightest sources (those brighter than $[3.6]=16$\,mag) are
indicated with the smaller hatched histogram.  Those of intermediate
brightness are indicated with the open histogram.
{\sl Right panel:}  As for the left panel, but for
SSDF sources selected at 4.5\,$\mu$m.
\label{fig:hist}
}
\end{figure}

The measured colors of celestial sources reflect their underlying nature, albeit 
after being folded through the detection/selection process.   
The IRAC colors for all sources listed in the two SSDF
catalogs are shown in Figure~\ref{fig:hist}.  
The color distributions are broadly consistent with what is seen
typically with IRAC at these flux levels, e.g., Ashby \etal\ (2009).
For example, the faintest SSDF sources (those fainter than 
[3.6]=[4.5]=17.5\,Vega mag) are systematically and significantly 
redder than brighter SSDF sources.  This is because
the surface density of relatively red extragalactic sources 
increases quickly below 17\,mag, while
the contribution from Galactic stars, which are relatively blue 
in the IRAC bands, flattens out (e.g., Fazio \etal 2004b).

The differential IRAC number counts in the SSDF are given in 
Table~\ref{tab:counts} and Figure~\ref{fig:counts}, 
after applying appropriate corrections for incompleteness 
based on the empirical estimates in Table~\ref{tab:comp}.  
Although the SSDF catalogs contain many sources 
brighter than 10\,mag, these are not shown in the source counts
because they are saturated in the IRAC mosaics.
Nonetheless the SSDF counts are 
broadly consistent with counts measured earlier by, e.g., 
Ashby \etal\ (2009) and Fazio \etal\ (2004b), 
over the magnitude range covered by the SSDF catalogs.
At bright flux levels, the SSDF counts are slightly elevated 
with respect to SDWFS.  We have examined the
SSDF mosaics at the locations of sources in the affected magnitude
ranges, and find that virtually all are pointlike.  This is 
consistent with a picture in which these sources are due to Milky Way
stars, seen along a line of sight that is closer in both 
latitude and longitude ($\ell,b) = (325,-58)$ to the Galactic
center than is SDWFS ($\ell,b) = (60,+67)$.  This is borne out
by the consistency between the DIRBE model Milky Way star counts
and the bright IRAC SSDF counts shown in Figure~\ref{fig:counts}.

In the 30\,s exposures used to construct the SSDF
mosaics, sources brighter than $\sim$10\,Vega mag in either IRAC
band
are saturated.  Users of the SSDF catalogs are cautioned
against uncritical usage of photometry for sources brighter
than [3.6]=[4.5]=11.5\,Vega mag.  
Objects that are truly as
bright as this will be well-detected in any case by the two 
short-wavelength {\sl WISE} bands.

\section{Summary}
\label{sec:summary}

We have carried out an infrared survey of nearly 100\,deg$^2$ with the
warm IRAC aboard \SSS\ for our Cycle 8 \SSS\ Exploration program, 
the \SSS-South Pole Telescope Deep Field survey.  With its
combination of uniform depth in two infrared bands and wide-area 
coverage, this project provides a unique resource for extragalactic research.  
It benefits from numerous coextensive observations spanning X-ray to 
radio wavelengths, in particular the deep imaging acquired by the 
South Pole Telescope at 1.4, 2, and 3\,mm, and will therefore
be of particular use for galaxy cluster science.  The catalogs
contain multiple photometric measurements for several million distinct 
IRAC sources down to the $5\sigma$ survey limits of 7.0 and 9.4\,$\mu$Jy at
3.6 and 4.5\,$\mu$m, respectively, and have been made 
publicly available to the astronomical community from the Spitzer Science Center.

\acknowledgments

\begin{figure}[htb]
\epsscale{1.0}
% \includegraphics[height=8in,width=7in,angle=0]{SSDF_counts.v9.pdf}
% \vskip-2in
\plotone{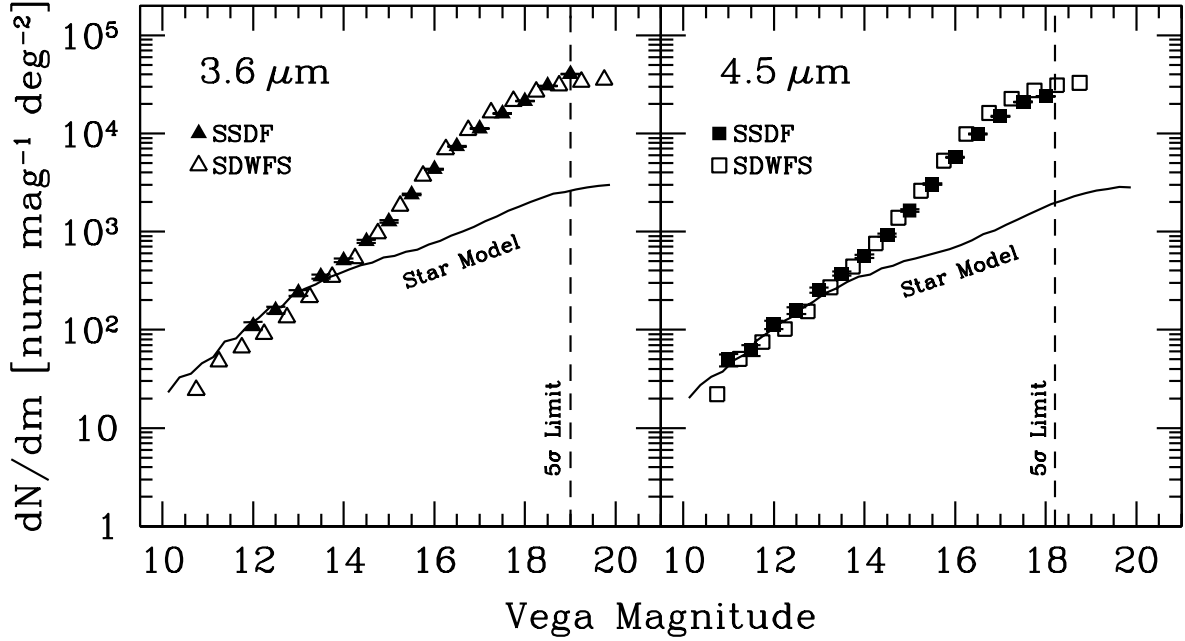}
\caption{
Differential source counts for the SSDF in the two IRAC bands.  The 3.6\,$\mu$m
counts (left panel, solid triangles) are taken from the 3.6\,$\mu$m-selected 
SSDF catalog, while the 4.5\,$\mu$m counts (right panel, solid squares) 
are taken from the 4.5\,$\mu$m-selected catalog.  All counts shown here are
based on SExtractor MAG\_AUTO estimates for unsaturated sources and are 
corrected for the effects of incompleteness using the empirical estimates
given in Table~\ref{tab:comp}.  The solid lines indicate the expected 
counts arising from Milky Way stars along the line of sight through the
center of the SSDF at ($\alpha,\delta)=($23:30,$-$55:00) based on the DIRBE
Faint Source Model at 3.5 and 4.9\,$\mu$m in the left and right 
panels, respectively (Arendt et al. 1998; Wainscoat et al. 1992; 
Cohen 1993; 1994; 1995).  The SSDF counts are similar to those
measured in the narrower but more sensitive \SSS\ Deep, Wide Field
Survey (open symbols; Ashby \etal\ 2009), as discussed in Section~\ref{sec:discussion}.
\label{fig:counts}
}
\end{figure}

This work is based on observations made with the {\it Spitzer Space
Telescope}, which is operated by the Jet Propulsion Laboratory,
California Institute of Technology under contract with the National 
Aeronautics and Space Administration (NASA).  Support was provided 
by NASA through contract number 1439357 issued by JPL/Caltech.  
IRAF is distributed by the National Optical Astronomy 
Observatory, which is operated by the Association of Universities for 
Research in Astronomy (AURA) under cooperative agreement with the 
National Science Foundation.  Lawrence Livermore National Laboratory
is operated by Lawrence Livermore National Security, LLC, for the
U.S.\ Department of Energy, National Nuclear Security Administration
under Contract DE-AC52-07NA27344.  FP acknowledges support from
grant 50 OR 1117 of the Deutches Zenturm f\"ur Luft- und Raumfahrt (DLR).
We thank Dave Nair for his efforts in characterizing a preliminary
reduction of the SSDF images.  We also thank Richard G. Arendt, 
who kindly computed the Milky Way star count models shown in 
Figure~\ref{fig:counts}.

Facilities:  \facility{{\it Spitzer Space Telescope} (IRAC)}

{}

\clearpage

% FIGURE 1 - etendue plot

% FIGURE 2 - the SSDF field boundaries and the BCS/XMM coverage

% FIGURE 3 - the SSDF field boundaries and the ATCA/SPIRE/SPT coverage

% FIGURE 4 - the SSDF tiling scheme and 4.5 um coverage map

% FIGURE 5 - the SSDF completeness at 4.5 um

% FIGURE - the SSDF positions relative to 2MASS

% FIGURE - the SSDF color distribution

% FIGURE - the SSDF source counts

% \begin{figure}[h]
% \includegraphics[trim=0mm 0mm 0mm
% 0mm,clip=True,width=\columnwidth]{aper_corr_both_channels.pdf}
% \caption{\label{aper_corr} Aperture corrections derived from the
% photometric simulation. The dashed lines represent the pre-correction
% derived from the cleanly-recovered sources, while the straight line is
% obtained from all recovered sources (see text for details). }
% \end{figure}

\clearpage

% TABLE -- the ancillary data
\begin{deluxetable}{ccrl}
\tabletypesize{\scriptsize}
\tablecaption{Available Survey Data in SSDF\label{tab:ancillary}}
\tablehead{
\colhead{Waveband}  & \colhead{Origin} & \colhead{Depth} \\
\colhead{($\mu$m)}  &                  & \colhead{(5$\sigma$)}
}
\startdata
0.5-0.9 ($ugriz$)   & Blanco Cosmology Survey & $\sim1.0\,\mu$Jy & Bleem et al., in preparation \\
1.35 ($J$)          & VISTA Hemisphere Survey & 12.0\,$\mu$Jy & McMahon et al., in preparation    \\
1.65 ($J$)          & VISTA Hemisphere Survey & 17.4\,$\mu$Jy & McMahon et al., in preparation    \\
2.20 ($K$)          & VISTA Hemisphere Survey & 30.2\,$\mu$Jy & McMahon et al., in preparation    \\
3.6                 & {\sl Spitzer}-SPT Deep Field & 7.0\,$\mu$Jy  & This work                    \\
4.5                 & {\sl Spitzer}-SPT Deep Field & 9.4\,$\mu$Jy  & This work                    \\
12                  & {\sl WISE W3}                & 1\,mJy & Wright et al. (2010)         \\
22                  & {\sl WISE W4}                & 6\,mJy & Wright et al. (2010)         \\
250                 & {\sl Hershel}/SPIRE          &10\,mJy & Holder et al. (2013)         \\
350                 & {\sl Hershel}/SPIRE          &10\,mJy & Holder et al. (2013)         \\
500                 & {\sl Hershel}/SPIRE          &10\,mJy & Holder et al. (2013)         \\
1400                & SPT                          &15\,mJy & Story et al. (2012)          \\
2000                & SPT                          & 5\,mJy & Story et al. (2012)          \\
3000                & SPT                          &10\,mJy & Story et al. (2012)          \\
\enddata
\tablecomments{Depths of coverage for other surveys of the SSDF.
}
\end{deluxetable}

% TABLE -- the tiling scheme
\begin{deluxetable}{ccl}
\tabletypesize{\scriptsize}
\tablecaption{SPT-Spitzer Tile Definitions\label{tab:tiles}}
\tablehead{
\colhead{Tile}  & \colhead{Right Ascension Range} & \colhead{Declination Range} \\
                     & \colhead{(Degrees, J2000)}      & \colhead{(Degrees, J2000)}
}
\startdata
SSDF0.0              & $\alpha \le 348.23$               & $-49.8 \ge \delta > -51.0$ \\
SSDF0.1              & $\alpha \le 348.00$               & $-51.0 \ge \delta > -52.0$ \\
SSDF0.2              & $\alpha \le 348.00$               & $-52.0 \ge \delta > -53.0$ \\
SSDF0.3              & $\alpha \le 348.00$               & $-53.0 \ge \delta > -54.0$ \\
SSDF0.4              & $\alpha \le 348.00$               & $-54.0 \ge \delta > -55.0$ \\
SSDF0.5              & $\alpha \le 348.00$               & $-55.0 \ge \delta > -56.0$ \\
SSDF0.6              & $\alpha \le 348.75$               & $-56.0 \ge \delta > -57.0$ \\
SSDF0.7              & $\alpha \le 348.75$               & $-57.0 \ge \delta > -58.0$ \\
SSDF0.8              & $\alpha \le 348.75$               & $-58.0 \ge \delta > -59.0$ \\
SSDF0.9              & $\alpha \le 348.75$               & $-59.0 \ge \delta > -60.2$ \\
SSDF1.0              & $348.23 <  \alpha \le 351.00$     & $-49.8 \ge \delta > -51.0$ \\
SSDF1.1              & $348.00 <  \alpha \le 351.00$     & $-51.0 \ge \delta > -52.0$ \\
SSDF1.2              & $348.00 <  \alpha \le 351.00$     & $-52.0 \ge \delta > -53.0$ \\
SSDF1.3              & $348.00 <  \alpha \le 351.00$     & $-53.0 \ge \delta > -54.0$ \\
SSDF1.4              & $348.00 <  \alpha \le 351.00$     & $-54.0 \ge \delta > -55.0$ \\
SSDF1.5              & $348.00 <  \alpha \le 351.00$     & $-55.0 \ge \delta > -56.0$ \\
SSDF1.6              & $348.75 <  \alpha \le 352.49$     & $-56.0 \ge \delta > -57.0$ \\
SSDF1.7              & $348.75 <  \alpha \le 352.45$     & $-57.0 \ge \delta > -58.0$ \\
SSDF1.8              & $348.75 <  \alpha \le 352.38$     & $-58.0 \ge \delta > -59.0$ \\
SSDF1.9              & $348.75 <  \alpha \le 352.35$     & $-59.0 \ge \delta > -60.2$ \\
SSDF2.0              & $351.00 <  \alpha \le 354.00$     & $-49.8 \ge \delta > -51.0$ \\
SSDF2.1              & $351.00 <  \alpha \le 354.00$     & $-51.0 \ge \delta > -52.0$ \\
SSDF2.2              & $351.00 <  \alpha \le 354.00$     & $-52.0 \ge \delta > -53.0$ \\
SSDF2.3              & $351.00 <  \alpha \le 354.00$     & $-53.0 \ge \delta > -54.0$ \\
SSDF2.4              & $351.00 <  \alpha \le 354.00$     & $-54.0 \ge \delta > -55.0$ \\
SSDF2.5              & $351.00 <  \alpha \le 354.00$     & $-55.0 \ge \delta > -56.0$ \\
SSDF2.6              & $352.49 <  \alpha \le 356.25$     & $-56.0 \ge \delta > -57.0$ \\
SSDF2.7              & $352.45 <  \alpha \le 356.25$     & $-57.0 \ge \delta > -58.0$ \\
SSDF2.8              & $352.38 <  \alpha \le 356.25$     & $-58.0 \ge \delta > -59.0$ \\
SSDF2.9              & $352.35 <  \alpha \le 356.25$     & $-59.0 \ge \delta > -60.2$ \\
SSDF3.0              & $354.00 <  \alpha \le 357.00$     & $-49.8 \ge \delta > -51.0$ \\
SSDF3.1              & $354.00 <  \alpha \le 357.00$     & $-51.0 \ge \delta > -52.0$ \\
SSDF3.2              & $354.00 <  \alpha \le 357.00$     & $-52.0 \ge \delta > -53.0$ \\
SSDF3.3              & $354.00 <  \alpha \le 357.00$     & $-53.0 \ge \delta > -54.0$ \\
SSDF3.4              & $354.00 <  \alpha \le 357.00$     & $-54.0 \ge \delta > -55.0$ \\
SSDF3.5              & $354.00 <  \alpha \le 357.00$     & $-55.0 \ge \delta > -56.0$ \\
SSDF3.6              & $356.25 <  \alpha; \alpha < 2.0$  & $-56.0 \ge \delta > -57.0$ \\
SSDF3.7              & $356.25 <  \alpha; \alpha < 2.0$  & $-57.0 \ge \delta > -58.0$ \\
SSDF3.8              & $356.25 <  \alpha; \alpha < 2.0$  & $-58.0 \ge \delta > -59.0$ \\
SSDF3.9              & $356.25 <  \alpha; \alpha < 2.0$  & $-59.0 \ge \delta > -61.2$ \\
SSDF4.0              & $357.00 <  \alpha; \alpha < 2.0$  & $-49.8 \ge \delta > -51.0$ \\
SSDF4.1              & $357.00 <  \alpha; \alpha < 2.0$  & $-51.0 \ge \delta > -52.0$ \\
SSDF4.2              & $357.00 <  \alpha; \alpha < 2.0$  & $-52.0 \ge \delta > -53.0$ \\
SSDF4.3              & $357.00 <  \alpha; \alpha < 2.0$  & $-53.0 \ge \delta > -54.0$ \\
SSDF4.4              & $357.00 <  \alpha; \alpha < 2.0$  & $-54.0 \ge \delta > -55.0$ \\
SSDF4.5              & $357.00 <  \alpha; \alpha < 2.0$  & $-55.0 \ge \delta > -56.0$ \\
\enddata
\tablecomments{The locations and dimensions of sub-regions (tiles) in which
the SSDF IRAC data were reduced in pixel-pixel registration.
}
\end{deluxetable}

% TABLE - SExtractor parameters
\begin{deluxetable}{lc}
\tabletypesize{\small}
\tabletypesize{\scriptsize}
\tablecaption{SSDF SExtractor Parameter Settings.\label{tab:settings}}
\tablewidth{0pt}
\tablehead{
\colhead{PARAMETER} & \colhead{SETTING} 
}
\startdata
DETECT\_MINAREA [pixel] & 4   \\
DETECT\_THRESH [sigma]  & 0.5 \\
FILTER & gauss\_3.0\_7x7 \\
DEBLEND\_NTHRESH & 64\\
DEBLEND\_MINCONT & 0.0001 \\
% SEEING\_FWHM [FWHM, arcsec] & 1.66 & 1.72 \\
BACK\_SIZE [pixel] & 128 \\
BACK\_FILTERSIZE & 3 \\
BACKPHOTO\_TYPE & GLOBAL \\
\enddata
\tablecomments{Parameter settings used to identify and photometer sources 
identically in both of the SSDF IRAC bands.  The only SExtractor
settings that differed in the two bands were SEEING\_FWHM and MAG\_ZERO.
SEEING\_FWHM was set to 1\farcs66 and 1\farcs72 in the 3.6 and 4.5\,$\mu$m 
mosaics, respectively.  MAG\_ZERO was set to 18.789 (3.6\,$\mu$m) 
and 18.316\,Vega mag (4.5\,$\mu$m).
}
\end{deluxetable}

% TABLE - aperture corrections
\begin{deluxetable}{ccc}
\tabletypesize{\scriptsize}
\tablecaption{SSDF Aperture Corrections.\label{tab:apcorr}}
\tablewidth{0pt}
\tablehead{
\colhead{Diameter}  & 3.6\,$\mu$m & 4.5\,$\mu$m \\
\colhead{(arcsec)}  & (mag)       & (mag)      
}
\startdata
2\arcsec  	&  $-$1.13  & $-$1.07  \\
3\arcsec  	&  $-$0.58  & $-$0.56  \\
4\arcsec  	&  $-$0.33  & $-$0.33  \\
5\arcsec  	&  $-$0.22  & $-$0.20  \\
6\arcsec  	&  $-$0.16  & $-$0.14  \\
7\arcsec  	&  $-$0.13  & $-$0.11  \\
8\arcsec  	&  $-$0.11  & $-$0.09  \\
9\arcsec  	&  $-$0.09  & $-$0.08  \\
10\arcsec  	&  $-$0.08  & $-$0.07  \\
15\arcsec  	&  $-$0.04  & $-$0.03  \\
\enddata
\tablecomments{Aperture corrections (magnitudes) 
derived from comparisons of photometry in the SSDF apertures to
that measured in the fiducial IRAC aperture (diameter $24^{\prime\prime}$).
These corrections are consistent with those tabulated in the IRAC Instrument
Handbook.
All photometry compiled in the SSDF catalogs presented in this work 
has been aperture-corrected to total magnitudes using these values.
}
\end{deluxetable}

% TABLE - completeness
\begin{deluxetable}{ccccc}
\tabletypesize{\scriptsize}
\tablecolumns{6}
\tablewidth{0pc}
\tablecaption{SSDF Completeness\label{tab:comp}}
\tablehead{
\colhead{Mag}  & \multicolumn{2}{c}{3.6\,$\mu$m} & \multicolumn{2}{c}{4.5\,$\mu$m} \\ 
\colhead{(Vega)} & Completeness & Unc. & Completeness & Unc. 
}
\startdata
% These are the old estimates, from the more aggressive
% recovery algorithm
% 15.0 & 0.996 & 0.996 \\
% 15.5 & 0.993 & 0.993 \\
% 16.0 & 0.988 & 0.986 \\
% 16.5 & 0.980 & 0.975 \\
% 17.0 & 0.965 & 0.953 \\
% 17.5 & 0.939 & 0.931 \\
% 18.0 & 0.906 & 0.879 \\
% 18.5 & 0.859 & 0.812 \\
% 19.0 & 0.760 & 0.697 \\
% 19.5 & 0.709 & 0.441 \\
% 20.0 & 0.539 & 0.231 \\
% 20.5 & 0.301 & 0.132 \\
% 21.0 & 0.155 & 0.080 \\
% These estimates are from the usual MC technique, but v9 not v8
 9.5 &\nodata&\nodata& 0.999 & 0.002 \\
10.0 & 0.998 & 0.002 & 0.998 & 0.002 \\
10.5 & 0.997 & 0.002 & 0.998 & 0.001 \\
11.0 & 0.997 & 0.001 & 0.997 & 0.003 \\
11.5 & 0.996 & 0.001 & 0.996 & 0.003 \\
12.0 & 0.993 & 0.003 & 0.995 & 0.002 \\
12.5 & 0.992 & 0.003 & 0.991 & 0.004 \\
13.0 & 0.988 & 0.005 & 0.990 & 0.003 \\
13.5 & 0.985 & 0.005 & 0.987 & 0.004 \\
14.0 & 0.982 & 0.006 & 0.981 & 0.004 \\
14.5 & 0.976 & 0.007 & 0.974 & 0.005 \\
15.0 & 0.968 & 0.006 & 0.968 & 0.002 \\
15.5 & 0.955 & 0.007 & 0.949 & 0.006 \\
16.0 & 0.933 & 0.007 & 0.932 & 0.005 \\
16.5 & 0.910 & 0.004 & 0.90  & 0.01  \\
17.0 & 0.88  & 0.01  & 0.84  & 0.01  \\
17.5 & 0.80  & 0.04  & 0.72  & 0.03  \\
18.0 & 0.71  & 0.03  & 0.55  & 0.03  \\
18.5 & 0.55  & 0.03  & 0.39  & 0.03  \\
19.0 & 0.39  & 0.03  & 0.27  & 0.02  \\
19.5 & 0.27  & 0.02  & 0.19  & 0.02  \\
20.0 & 0.18  & 0.01  & 0.11  & 0.01  \\
20.5 & 0.10  & 0.01  &\nodata&\nodata\\
\enddata
\tablecomments{Completeness estimates and uncertainties for the SSDF 
at 3.6 and 4.5\,$\mu$m.  Uncertainties are empirical estimates based 
on variations in completeness measured in five separate tiles.
The completeness is unity at magnitudes brighter than those listed,
although such sources will be saturated in the SSDF mosaics.
}
\end{deluxetable}

\begin{deluxetable}{rlll}
\tabletypesize{\scriptsize}
\tablecaption{SPT-Spitzer Photometry Catalog Column Definitions\label{tab:catalog}}
\tablehead{
\colhead{Column}  & \colhead{Parameter} & \colhead{Description} & \colhead{Units} 
}
\startdata
  1 & TILE             & SSDF sub-tile of origin                           &       \\
  2 & X\_IMAGE         & Object position along x                           & pixel \\
  3 & Y\_IMAGE         & Object position along y                           & pixel \\
  4 & ALPHA\_J2000     & Right ascension of barycenter (J2000)             & deg   \\
  5 & DELTA\_J2000     & Declination of barycenter (J2000)                 & deg   \\
  6 & KRON\_RADIUS     & Kron apertures in units of A or B                 &       \\
  7 & BACKGROUND       & Background at centroid position                   & count \\
  8 & FLUX\_RADIUS     & Fraction-of-light radii                           & pixel \\
  9 & ALPHAPEAK\_J2000 & Right ascension of brightest pix (J2000)          & deg   \\
 10 & DELTAPEAK\_J2000 & Declination of brightest pix (J2000)              & deg   \\
 11 & X2\_IMAGE        & Variance along x                                  & pixel$^2$ \\
 12 & Y2\_IMAGE        & Variance along y                                  & pixel$^2$ \\
 13 & XY\_IMAGE        & Covariance between x and y                        & pixel$^2$ \\
 14 & A\_IMAGE         & Profile RMS along major axis                      & pixel \\
 15 & B\_IMAGE         & Profile RMS along minor axis                      & pixel \\
 16 & THETA\_IMAGE     & Position angle (CCW/x)                            & deg   \\
 17 & A\_WORLD         & Profile RMS along major axis (world units)        & deg   \\
 18 & B\_WORLD         & Profile RMS along minor axis (world units)        & deg   \\
 19 & THETA\_WORLD     & Position angle (CCW/world-x)                      & deg   \\
 20 & CLASS\_STAR      & S/G classifier output                             &       \\
 21 & FLAGS            & SExtractor flags                                  &       \\
\cutinhead{The following 12 quantities correspond to IRAC 3.6\,$\mu$m measurements} 
 22 & MAG\_AUTO        & Kron-like elliptical aperture magnitude           & Vega mag \\
 23 & MAGERR\_AUTO     & RMS error for AUTO magnitude                      & Vega mag \\
 24 & MAG\_APER        & 4\arcsec\ Diameter Aperture Magnitude, corrected  & Vega mag \\
 25 & MAGERR\_APER     & 4\arcsec\ Diameter Aperture Magnitude Uncertainty & Vega mag \\
 26 & MAG\_APER        & 6\arcsec\ Diameter Aperture Magnitude, corrected  & Vega mag \\
 27 & MAGERR\_APER     & 6\arcsec\ Diameter Aperture Magnitude Uncertainty & Vega mag \\
 28 & FLUX\_AUTO       & Kron-like elliptical aperture flux                & $\mu$Jy \\
 29 & FLUXERR\_AUTO    & RMS error for AUTO flux                           & $\mu$Jy \\
 30 & FLUX\_APER       & 4\arcsec\ Diameter Aperture Flux, corrected       & $\mu$Jy \\
 31 & FLUXERR\_APER    & 4\arcsec\ Diameter Aperture Flux, Uncertainty     & $\mu$Jy \\
 32 & FLUX\_APER       & 6\arcsec\ Diameter Aperture Flux, corrected       & $\mu$Jy \\
 33 & FLUXERR\_APER    & 6\arcsec\ Diameter Aperture Flux, Uncertainty     & $\mu$Jy \\
\cutinhead{The following 12 quantities correspond to IRAC 4.5\,$\mu$m measurements}
 34 & MAG\_AUTO        & Kron-like elliptical aperture magnitude           & Vega mag \\
 35 & MAGERR\_AUTO     & RMS error for AUTO magnitude                      & Vega mag \\
 36 & MAG\_APER        & 4\arcsec\ Diameter Aperture Magnitude, corrected  & Vega mag \\
 37 & MAGERR\_APER     & 4\arcsec\ Diameter Aperture Magnitude Uncertainty & Vega mag \\
 38 & MAG\_APER        & 6\arcsec\ Diameter Aperture Magnitude, corrected  & Vega mag \\
 39 & MAGERR\_APER     & 6\arcsec\ Diameter Aperture Magnitude Uncertainty & Vega mag \\
 40 & FLUX\_AUTO       & Kron-like elliptical aperture magnitude           & $\mu$Jy \\
 41 & FLUXERR\_AUTO    & RMS error for AUTO magnitude                      & $\mu$Jy \\
 42 & FLUX\_APER       & 4\arcsec\ Diameter Aperture Flux, corrected       & $\mu$Jy \\
 43 & FLUXERR\_APER    & 4\arcsec\ Diameter Aperture Flux Uncertainty      & $\mu$Jy \\
 44 & FLUX\_APER       & 6\arcsec\ Diameter Aperture Flux, corrected       & $\mu$Jy \\
 45 & FLUXERR\_APER    & 6\arcsec\ Diameter Aperture Flux Uncertainty      & $\mu$Jy \\
\enddata
\tablecomments{The column definitions for both \SSS-SPT Deep Field
catalogs.  All columns except column 1 contain quantities output by SExtractor
(Bertin \& Arnouts 1996); column 1 specifies the mosaic sub-fields (tiles; see
Figure~\ref{fig:field} and Table~\ref{tab:tiles}) processed individually by SExtractor to generate the photometric measurements tabulated in these catalogs.
This table is available in its entirety in machine-readable form in the online
journal.  
}
\end{deluxetable}

\begin{deluxetable}{r cc cc }
\tabletypesize{\small}
\tabletypesize{\scriptsize}
\tablecolumns{5}
\tablewidth{0pc}
\tablecaption{SSDF IRAC Number Counts\label{tab:counts}}
\tablehead{
\colhead{Mag}  & \multicolumn{2}{c}{3.6\,$\mu$m} & \multicolumn{2}{c}{4.5\,$\mu$m} \\ 
\colhead{(Vega)} & Counts & Unc. & Counts & Unc. 
}
\startdata
11.0  &  \nodata & \nodata &  1.69 & 0.058 \\ 
11.5  &  \nodata & \nodata &  1.79 & 0.052 \\ 
12.0  &  2.04 & 0.040 &  2.05 & 0.039 \\ 
12.5  &  2.20 & 0.033 &  2.20 & 0.033 \\ 
13.0  &  2.38 & 0.027 &  2.40 & 0.026 \\ 
13.5  &  2.54 & 0.023 &  2.57 & 0.022 \\ 
14.0  &  2.71 & 0.019 &  2.75 & 0.018 \\ 
14.5  &  2.90 & 0.015 &  2.97 & 0.014 \\ 
15.0  &  3.10 & 0.012 &  3.21 & 0.011 \\ 
15.5  &  3.38 & 0.009 &  3.49 & 0.008 \\ 
16.0  &  3.63 & 0.007 &  3.76 & 0.006 \\ 
16.5  &  3.87 & 0.005 &  4.00 & 0.004 \\ 
17.0  &  4.05 & 0.004 &  4.17 & 0.004 \\ 
17.5  &  4.20 & 0.003 &  4.32 & 0.003 \\ 
18.0  &  4.33 & 0.003 &  4.38 & 0.003 \\ 
18.5  &  4.49 & 0.002 &  \nodata & \nodata \\ 
19.0  &  4.61 & 0.002 &  \nodata & \nodata \\ 
\enddata
\tablecomments{Differential SSDF number counts measured in bins of width
0.5\,mag centered at the magnitudes given in the left-hand column.  Counts
are expressed in terms of log$(N)$\,mag$^{-1}$\,deg$^{-2}$.  
All uncertainties are 1$\sigma$ and reflect Poisson counting statistics 
only; uncertainties arising from the incompleteness correction will dominate
at faint levels.  
}
\end{deluxetable}

%%
%% End of file 
\end{document}